\documentclass[epj,twocolumn]{webofc}
\usepackage[varg]{txfonts}   
\usepackage{graphicx}

\def\apj{ApJ\,}

\def\aap{A\&A\,}

\def\mnras{MNRAS\,}

\def\nat{Nature}
\def\apjs{ApJS}

\wocname{epj}
\woctitle{Seismology of the Sun and the Distant Stars 2016}
\begin{document}
\title{Evolutionary states of red-giant stars from grid-based modelling}
%

\author{\firstname{Saskia} \lastname{Hekker}\inst{1,2}\fnsep\thanks{\email{Hekker@mps.mpg.de}} \and
        \firstname{Yvonne} \lastname{Elsworth}\inst{3,2}\and
        \firstname{Sarbani} \lastname{Basu}\inst{4}\and
        \firstname{Earl} \lastname{Bellinger}\inst{1,2,4}        
}

\institute{Max Planck Institute for Solar System Research, G\"ottingen, Germany
\and
           Stellar Astrophysics Centre, Aarhus, Denmark 
\and
           University of Birmingham, Birmingham, United Kingdom
\and
           Yale University, New Haven, United States of America
          }

\abstract{%
From its surface properties it can be difficult to determine whether a red-giant star is in its helium-core-burning phase or only burning hydrogen in a shell around an inert helium core. Stars in either of these stages can have similar effective temperatures, radii and hence luminosities, i.e. they can be located at the same position in the Hertzsprung-Russell diagram. Asteroseismology -- the study of the internal structure of stars through their global oscillations -- can provide the necessary additional constraints to determine the evolutionary states of red-giant stars. Here, we present a method that uses grid-based modelling based on global asteroseismic properties ($\nu_{\rm max}$, frequency of maximum oscillation power; and $\Delta\nu$, frequency spacing between modes of the same degree and consecutive radial orders) as well as effective temperature and metallicity to determine the evolutionary phases. This method is applicable even to timeseries data of limited length, although with a small fraction of miss-classifications. 
}
\maketitle
%
\section{Introduction}
\label{intro}
The long near-uninterrupted photometric timeseries of \textit{Kepler} data have unambiguously shown the presence of mixed pressure-gravity oscillation modes. These modes provide information about the stellar structure of red giants and from that allow to identify the evolutionary state of a red giant \cite{bedding2011,mosser2014}, i.e. whether the star is on the red-giant branch (RGB) or in the He-core burning phase. However, resolving individual mixed modes is not always feasible in shorter timeseries data such as those of the K2 \cite{haas2014} and TESS \cite{ricker2015} missions.
Here we present a method that does not rely on the detection of the individual mixed modes, but uses grid-based modelling (GBM).

GBM is the technique in which a set of observed parameters of a star are compared to these parameters obtained from a grid of stellar models to infer in this way the most likely stellar parameters of the observed star.  In asteroseismic GBM of red giants the observed parameters are generally $\nu_{\rm max}$ (frequency of maximum oscillation power), $\Delta\nu$ (frequency spacing between modes of the same degree and consecutive radial orders), effective temperature and metallicity.

Within the asteroseismology community there is a general thought that for GBM applied to red-giant stars it is important to know a priori the evolutionary phase, i.e., whether a star is in the hydrogen-shell burning red-giant branch (RGB)  phase or in the He-core burning phase. This knowledge is important to avoid biases in the resulting stellar parameters, i.e., mass, radius, age etc. Here, we follow this reasoning in the opposite direction and seek to answer the question whether it is possible to determine the evolutionary phases of red-giant stars from grid-based modelling using as input $\Delta\nu$, $\nu_{\rm max}$, $T_{\rm eff}$ and [Fe/H]. In other words, we want to determine whether or not it is possible to find loci in this four dimensional space where RGB or He-core burning stars are uniquely defined, thereby allowing us to provide a reliable identification of the evolutionary phase.

An earlier version of the method presented here has been applied together with the methods by \cite{kallinger2012,mosser2014,elsworth2016} to the APOKASC stars \cite[][see also the contributions by Marc Pinsonnealt and Yvonne Elsworth in these proceedings]{pinsonneault2014} to provide consolidated evolutionary phases.

\section{Methodology}
\label{sec-methodology}
\begin{figure*}
\centering
\includegraphics[width=0.9\linewidth]{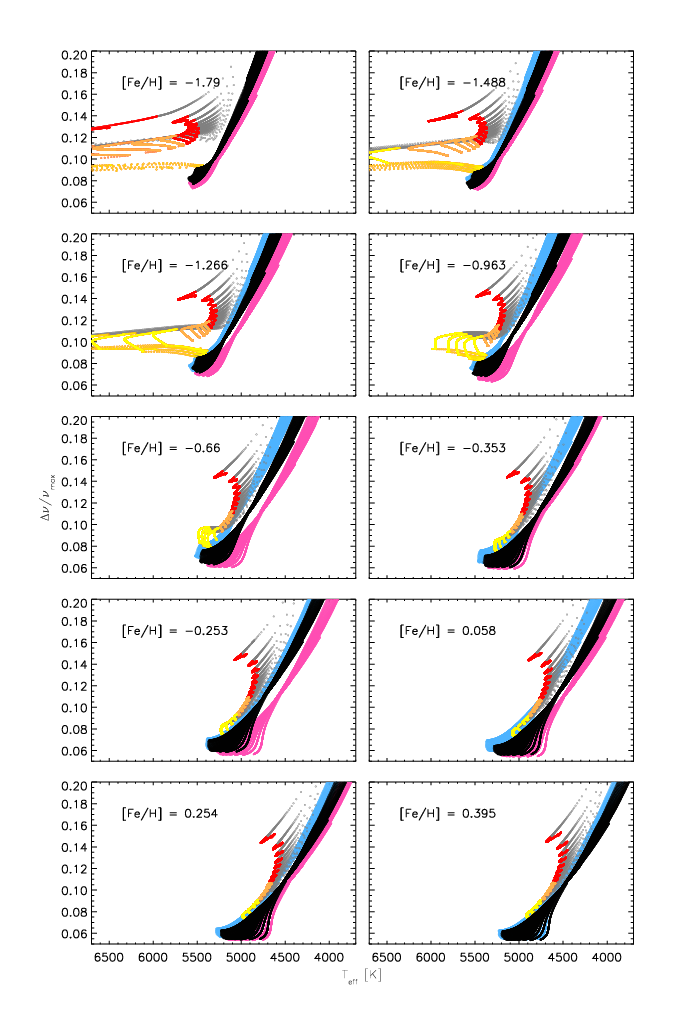}
\caption{$\Delta\nu/\nu_{\rm max}$ as a function of $T_{\rm eff}$ for different [Fe/H] values as indicated in the legend of each panel for BASTI models. Models of RGB stars are indicated in black. He-burning stars with He core mass fraction $>$ 0.1 are indicated in red, orange and yellow for models with masses $<$ 1.5~M$_{\odot}$, masses between 1.5 and 2.0 M$_{\odot}$ and masses 2.0 and 2.5 M$_{\odot}$, respectively. He-core burning stars with He core mass fraction between 0.0 and 0.1 are shown in gray. The models in light blue (pink) indicate RGB models with lower (higher) metallicities in the grid. In this figure we have defined RGB stars as those stars that lie between the base and the tip of the red-giant branch as defined by evolutionary key points 5 and 8 in the BaSTI documentation.}
\label{dnunumaxfig}       
\end{figure*}

By using GBM to find the evolutionary phase we are searching in a four-dimensional ($\Delta\nu$, $\nu_{\rm max}$, $T_{\rm eff}$, [Fe/H]) space for loci that uniquely define the evolutionary phase. To do so we have used a grid of BASTI models \cite{pietrinferni2004} to which we applied a Support Vector Machine (SVM). We did this in the four-dimensional space described above as well as in a three-dimensional space where we used the ratio  $\Delta\nu / \nu_{\rm max}$ together with $T_{\rm eff}$, [Fe/H]. The distribution of the models in this reduced parameter space is shown in figure~\ref{dnunumaxfig}. More details of the models and the SVM are discussed below.

\subsection{BaSTI models}
\label{sec-basti}
The models are from the canonical BaSTI grid\footnote{http://albione.oa-teramo.inaf.it/} \cite{pietrinferni2004} using an updated version of the code described by \cite{cassisi1997} and \cite{salaris1998}. The grid spans masses from 0.7~M$_{\odot}$ to 2.5~M$_{\odot}$ in steps of 0.1~M$_{\odot}$ and metallicities of $Z=0.0003$, $0.0006$, $0.001$, $0.002$, $0.004$, $0.008$, $0.01$, $0.0198$, $0.03$ and $0.04$. 

To compute $\Delta\nu$ and $\nu_{\rm max}$ for the models we use the scaling relations \cite{kjeldsen1995}:
\begin{equation}
\Delta\nu \propto \sqrt{\frac{M}{R^3}}
\label{dnu}
\end{equation}
\begin{equation}
\nu_{\rm max} \propto \frac{M}{R^2\sqrt{T_{\rm eff}}}
\label{numax}
\end{equation}
for which we used the following values as references: $\nu_{\rm max,ref} = \nu_{{\rm max},\odot}=3090\,\mu$Hz \cite{hekker2013}, and $\Delta\nu_{\rm ref}$ computed by \cite{guggenberger2016} to mitigate the known temperature and metallicity dependence.

\subsection{Support Vector Machine}
\label{sec-svm}
To find loci that uniquely define the evolutionary phases of red giants we train a support vector machine (SVM) for classification. SVMs are supervised learning models that are based on the concept of decision planes that define decision boundaries, with support vectors being the training data that are closest to the decision boundary. A decision plane is a plane that separates a set of objects having different class memberships. The basic idea behind SVM is to map the original objects using a set of mathematical functions, known as kernels, from the real space to a high-dimensional feature space in which classes can be separated. In this case we used a Gaussian radial basis function as a kernel:
\begin{equation}
\phi(r) = e^{-(\epsilon r)^2}
\label{rbf}
\end{equation}
with $r=||x_j - x_i||$ being the distance between the input values of $x_j$ and $x_i$, and $\epsilon$ is an adjustable parameter. 

In addition to the kernels one can choose the type of SVM. We used a C-type SVM in which training involves minimisation of the following error function:
\begin{equation}
\frac{1}{2}w^Tw+c\Sigma_{i=1}^N\xi_i
\label{minim}
\end{equation}
subject to the constraints:
\begin{equation}
y_i(w^T\phi(x_i)+b)\ge1-\xi_i
\label{constraints}
\end{equation}
in which $\xi_i \ge 0,\, i=1,... ,N$, $c$ is the capacity constant, $w$ is the vector of coefficients, $b$ is a constant, and $\xi_i$ represents parameters for handling non-separable input data, i.e. training data that cannot be classified correctly. The index $i$ labels the $N$ training cases.

A SVM searches for a hyperplane with the largest minimum margin separating the two classified groups. At the same time a SVM searches for a hyperplane that correctly separates as many instances as possible. The problem is that these requirements cannot always be accomplished at the same time. The capacity constant determines the balance between these requirements, with a low value of $c$ favouring towards a larger minimum margin and a high value of $c$ favouring the largest number of correct classifications in the training data.

For more details about SVM we refer to the original SVM paper \cite{vapnik1964} and the paper that made them computationally viable \cite{cortes1995}. Additionally, both radial basis functions and SVM are covered in detail in \cite{friedman2001}.

\section{Results}
\label{sec-res}
For our current set-up we used the R \cite{Rmanual} library kernlab \cite{karatzoglou2004} to train a C-type classification SVM, with $c=1$ and the a Gaussian radial base kernel as described above. In this set-up we trained models with masses in the range $0.75$~M$_{\odot} \leq M \leq 1.75$~M$_{\odot}$ and used as input parameters $T_{\rm eff}$, $\Delta\nu / \nu_{\rm max}$ and [Fe/H] (see Figure 1). For these models we can show that the RGB and He-core burning stars are indeed uniquely defined, i.e., we find a training error of 0 in the ideal case without uncertainties. We repeated the same experiment using a four-dimensional parameters space with $\Delta\nu$, $\nu_{\rm max}$, $T_{\rm eff}$, [Fe/H] as input data. In this case we could not reduce the training error to 0. This indicates that the four-dimensional input space is too complex to find a decision boundary that correctly separates all input data, while this is possible in the reduced three-dimensional parameter space.

\section{Prospects}
\label{sec-fut}
The determination of the evolutionary phase of red-giant stars based on grid-based modelling seems very promising.
A preliminary estimate for the success rate of this method is on the order of 80 to 90\%. As this method is applicable to relatively short asteroseismic timeseries, i.e., 30 to 80 days, as obtained with the TESS and K2 missions for which only $\Delta\nu$ and $\nu_{\rm max}$ can be extracted, this can be of significant added value.

The preliminary analysis presented here will need to undergo further tests to investigate the importance of observational uncertainties on the classification as well as the impact of different stellar evolution models. Results of the full series of tests and validations as well as success rate of this method will be published in a forthcoming paper.
%
%

\section*{Acknowledgements}
The research leading to the presented results has received funding from the European Research Council under the European Community's Seventh Framework Programme (FP7/2007-2013) / ERC grant agreement no 338251 (StellarAges). YE acknowledge the support of the UK Science and Technology Facilities Council (STFC). SB acknowledges support from NSF grant AST-1514676 and NASA grant NNX16AI09G. Funding for the Stellar Astrophysics Centre (SAC) is provided by The Danish National Research Foundation (Grant agreement no.: DNRF106).
This work has made use of BaSTI web tools.

\end{document}